\documentclass[10pt, letterpaper, conference]{IEEEtran}
\IEEEoverridecommandlockouts
\usepackage{cite}
\usepackage{amsmath,amssymb,amsfonts}
\usepackage{algpseudocode}
\usepackage{booktabs}
\usepackage[ruled, vlined, norelsize, linesnumbered]{algorithm2e}
\usepackage{algorithmicx}
\usepackage{graphicx}
\usepackage{textcomp}
\usepackage{comment}
\usepackage{xcolor}
\usepackage{soul}
\usepackage[bookmarks=false,hidelinks]{hyperref}
\SetKwInput{KwInput}{Input}
\SetKwInput{KwOutput}{Output}

\usepackage{multirow}

\usepackage{acronym}

\usepackage{balance}

\begin{document}

\title{Lightweight Countermeasures Against \\
Static Power Side-Channel Attacks}

%\author{}
\author{Jitendra Bhandari, 
Mohammed Nabeel,
Likhitha Mankali, 
Ozgur Sinanoglu,~\IEEEmembership{Senior Member,~IEEE},\\
Ramesh Karri,~\IEEEmembership{Fellow,~IEEE}, 
and Johann Knechtel,~\IEEEmembership{Member,~IEEE}
\thanks{This work was supported in part by the NYU Center for Cybersecurity (CCS) and the NYUAD CCS.%
	%Ganesh Gore, Xifan Tang, Pierre-Emmanuel Gaillardon are supported by AFRL and DARPA under agreement number FA8650-18-2-7855, and Scott Temple, Pierre-Emmanuel Gaillardon are supported by AFRL and DARPA under agreement number FA8650-18-2-7849.
	\protect\\ 
% \indent J. Bhandari and A. Khader Thalakkattu Moosa contributed equally to this work.\protect\\
\indent J.~Bhandari, L.~Mankali, and R. Karri are with New York University, New York City, NY, 11201 USA. E-mail: \{jb7410, lm4344, rkarri\}@nyu.edu\protect\\
\indent J.~Knechtel, M.~Nabeel, and O.~Sinanoglu are with New York University Abu Dhabi, UAE. E-mail: \{johann, mtn2, ozgursin\}@nyu.edu\protect
}}

\date{November 2019}
% \footnote{978-1-6654-3274-0/21/$31.00 ©2021 IEEE}
\maketitle

\IEEEpubidadjcol

\begin{abstract}
This paper presents a novel defense strategy against static power side-channel attacks (PSCAs), a critical threat to cryptographic security. Our method is based on
(1) carefully tuning high-Vth versus low-Vth cell selection during synthesis, accounting for both security and timing impact,
and (2), at runtime, randomly switching the operation between these cells. This approach serves to significantly obscure static power patterns,
which are at the heart of static PSCAs.
Our experimental results on a commercial 28nm node show a drastic increase in the effort required for a successful attack, namely up to 96 times more traces.
When compared to prior countermeasures, ours incurs little cost, making it a lightweight defense.

\end{abstract}

\begin{IEEEkeywords}
Power Side-Channel, Countermeasure
\end{IEEEkeywords}

\section{Introduction}

% With the rapid expansion of technological capabilities aimed at addressing increasing demands, we witness the emergence of sub-nanometer transistors. While this innovation plays a crucial role in tackling the need for fast integrated circuits (ICs), it concurrently introduces a set of challenges that necessitate careful consideration from a design perspective. Notably, these advancements escalate security vulnerabilities, as potential attackers can now access more substantial information leakage than what was possible with older technologies.

% More specifically, to facilitate quicker transitions, there is a requirement to lower the threshold voltage (Vt).
% Within any commercial library, one can observe a spectrum of cells characterized by diverse Vt properties.
% These range from fast-acting cells to low-power variants, the latter being particularly crucial for devices reliant on battery power.
% An example is shown in \autoref{tab:DFF_leak} for a D flip-flop (D-FF) across three different cell types, namely LVT (low Vt), RVT (regular Vt), and HVT (high Vt), respectively.
% There is a few magnitudes increase in leakage power as one transitions from HVT cells to LVT cells.

With the rapid advancement of technology to meet growing demands, sub-nanometer transistors have emerged, significantly enhancing the speed of integrated circuits (ICs). However, this innovation also introduces design challenges, particularly increasing security vulnerabilities due to greater information leakage.
To achieve faster transitions, there is a need to lower the threshold voltage (Vt). Commercial libraries offer a range of cells with varying Vt properties, from fast-acting to low-power cells, which are vital for battery-powered devices. An example is shown in \autoref{tab:DFF_leak} for a D flip-flop (D-FF) across three different cell types, namely LVT (low Vt), RVT (regular Vt), and HVT (high Vt), respectively.
There is a few magnitudes increase in leakage power as one transitions from HVT cells to LVT cells.

% From a designer's standpoint, the objective is to limit the utilization of LVT cells as much as possible, aiming to curtail power consumption. Still, due to timing failures in certain pathways, the use of
% these high-speed cells becomes an unavoidable necessity in specific contexts, timing closure in particular. If such modifications are done in traditional CAD flows, i.e., without security in mind,      the implications for the resilience of the circuits can be  negative~\cite{leakage-2010-TCAS,amstatic2014}.
% At the same time, as we show in this work, a careful and security-aware tuning of LVT cells usage can enable an competitive yet lightweight defense mechanism.

% Side-channel attacks present a significant threat to security, especially as they have proven to be effective against cryptographic algorithms, compromising their security assurances due to hardware limitations. A
% variety of side channels, including power, timing, and electromagnetic, have been explored extensively~\cite{EM-radiation,DPA,kocher1996timing}, revealing their potential to leak confidential information. Among these, the power side-channel has been the focus
% of most attacks~\cite{survey-PSC}, particularly for the dynamic power side-channel (D-PSC). This is because dynamic power contributes the significant share of total power, especially for older technologies.
% For modern and smaller technology nodes, however, especially with the use of low Vt cells, leakage power effects become significant and noticeable.
% Thus, static power side-channel (S-PSC) attacks become more promising as well.
Designers aim to limit LVT cell use to reduce power consumption, but high-speed cells are sometimes necessary for timing closure. Without security considerations, traditional CAD modifications can weaken circuit resilience~\cite{leakage-2010-TCAS,amstatic2014}. This work shows that careful, security-aware tuning of LVT cells can provide effective, lightweight defense.
Side-channel attacks, particularly against cryptographic algorithms, exploit hardware limitations and have been extensively studied~\cite{DPA,kocher1996timing}. Power side-channel attacks are most common~\cite{survey-PSC}, especially dynamic power side-channel (D-PSC) attacks, due to dynamic power's significant share in older technologies. In modern nodes with low Vt cells, leakage power is more significant, making static power side-channel (S-PSC) attacks increasingly relevant.

Our contributions are threefold: 
\begin{enumerate}
\item A lightweight countermeasure against S-PSC attacks.
\item Use of different Vt and drive strengths as a defense.
\item Mixing strategies yielding a low-overhead, resilient implementation. 
\end{enumerate}

% \textcolor{red}{Clearly mention measurements as future work; whereas presented results are best-case scenario as tech-accurate libs and simulation w/o any noises
% Rephrase masking Drop direct comparison ( Review, include dynamic PSCA )}

\begin{table}[t]
\caption{Normalized Leakage Currents for a D-FF in a Commercial 28nm Technology}

\begin{center}
\footnotesize
\begin{tabular}{|c|c|c|c|c|c|c|c|c|c|c|c|}
\hline

\multirow{2}{*}{CLK} & \multirow{2}{*}{D} & \multirow{2}{*}{Q} & \multicolumn{3}{c|}{{Leakage Current [Norm.]}}     \\
\cline{4-6}
 & & & LVT & RVT & HVT\\ \hline

0 & 0 & 0 & 112.8$\times$ & 9.0$\times$ & 1$\times$ \\ \hline
0 & 0 & 1 & 136.0$\times$ & 10.1$\times$ & 1$\times$ \\ \hline
0 & 1 & 0 & 129.3$\times$ & 10.1$\times$ & 1$\times$  \\ \hline
0 & 1 & 1 & 118.3$\times$ & 9.2$\times$ & 1$\times$  \\ \hline
1 & 0 & 0 & 138.1$\times$ & 10.2$\times$ & 1$\times$ \\ \hline
1 & 0 & 1 & 125.0$\times$ & 9.1$\times$ & 1$\times$ \\ \hline
1 & 1 & 0 & 131.5$\times$ & 9.7$\times$ & 1$\times$ \\ \hline
1 & 1 & 1 & 93.5$\times$ & 7.1$\times$ & 1$\times$  \\ \hline

\end{tabular}
\label{tab:DFF_leak}
\end{center}
\vspace{-2em}
\end{table}

\section{Background and Motivation\label{sec:background}}
Power side-channel  attacks (PSCA) measure the power consumption of a device to extract sensitive data like secret keys in cryptographic chips. There are two variants of the threat model: with or without
controlling the input texts. We adopt the more stringent latter variant.
We assume the attacker has control over the clock, which is common for PSCAs.
Prominent options for PSCAs include differential power analysis (DPA)~\cite{kocher1996timing}, and correlation power analysis (CPA)~\cite{brier2004CPA,knechtel22_PSC}. DPA compares power consumed by similar operations to retrieve secret key, whereas CPA uses the Pearson correlation coefficient to correlate the actual and predicted power profiles to infer the secret key. We employ CPA.

PSCAs can be conducted as attacks on dynamic (D-PSCAs) or static power (S-PSCAs).
A D-PSCA extracts secret data by observing and interpreting power use while a device is active, whereas a S-PSCA does so when the device is idle.
Note that data-related power patterns are strongly expressed even in the absence of active data processing (\autoref{tab:DFF_leak}).
While D-PSCAs require accurate timing, {S-PSCAs work with halted clocks, rendering this attack easier to conduct in the real world.}

Prior work has used various methods including masking to hide the unique power profiles of circuits during sensitive computations.
While prior work on D-PSC has shown effectiveness of these techniques, it incurs overhead where the final design can be multiple times larger than the original. This makes it impractical in the real-world where cost per mm\textsuperscript{2} of silicon is high. 
Further, most of these countermeasures are tailored against D-PSCAs. There is a subtle difference when it comes to countermeasures that aims to thwart the S-PSC information leakage:
irrespective of the additional circuitry used for masking, the parts that store the actual data can still leak information \cite{our_paper1,10.1007/978-3-319-57339-7_5,cryptography5030016}.
In other words, regular countermeasures  schemes devised against D-PSCAs do not directly apply against S-PSCAs.

\section{Related Work\label{sec:related}}

The work in \cite{10.1145/1228784.1228808} unveiled the S-PSC's potential as a security vulnerability. Practical experimentation with S-PSCAs utilizing FPGAs was later undertaken in \cite{amstatic2014}, marking one
of the initial research in this domain. The significant impact of leakage power on PSC, particularly in advanced technology nodes, was underscored in \cite{leakage-2010-TCAS}. In a similar direction, \cite{8888247}
provided an experimental analysis of how various measurement factors influence the success rate of S-PSCAs. The work of \cite{Karimi_Moos_Moradi_2019} highlighted the increasing effects of aging on smaller
technology nodes, thus escalating the security risks of contemporary devices when subjected to S-PSCAs. A comprehensive multivariate analysis focusing on S-PSC was presented in \cite{10.1007/978-3-319-57339-7_5}, while \cite{DATE-SC-Static} delved into both static and dynamic PSC in the context of the 65nm technology node, demonstrating the vulnerability of even older nodes to S-PSCAs.
The work in \cite{our_paper1} studied the role of Vt cells on the S-PSC from the perspective of hardware Trojan attacks.

\cite{Moos_Moradi_2021} provided a thorough evaluation of countermeasures against S-PSCAs, including logic balancing, noting the substantial overheads associated with various techniques.
Their analysis, conducted on a 28nm IC, also highlighted the critical role of different Vt cells.
\cite{9040870} introduced  standard-cell delay-based dual-rail pre-charge logic (SC-DDPL) as a 
countermeasure against S-PSC. The scheme uses NAND gates to enhance design symmetry and power profile uniformity. This method is incompatible with commercial CAD tool optimization flows, a notable limitation.
Masking schemes like \cite{8288553,QuadSeal} offer promising resilience, albeit at the cost of high overheads.
Furthermore, machine learning has been utilized to tune synthesis \cite{ASCENT} or runtime operation \cite{ML-assisted_countermeasure} toward higher resilience.

\section{Methodology \label{sec:framework}}
Our work is based on the observation that
the type of Vt cells substantially impacts the S-PSC, with orders of magnitudes shown in \autoref{tab:DFF_leak}.
\begin{figure}[t]
    \centering
    \includegraphics[width=0.65\columnwidth]{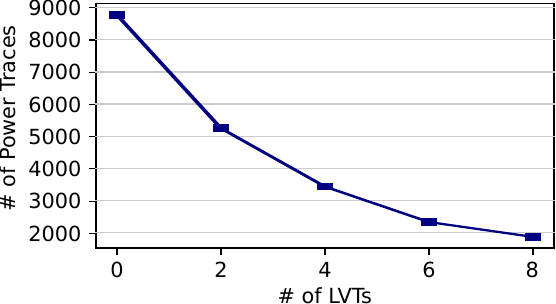}
    \caption{Baseline AES design under CPA attack. Shown are average numbers of power traces required until disclosure with 90\% success rate across a varying number of LVT cells (versus HVT and RVT cells) employed in the state registers that are grouped into bytes.
    }
    \label{fig:base}
\end{figure}
For common crypto cores like AES, state registers which hold the intermediate cipher values after each round are the most vulnerable among all the FFs \cite{our_paper1,Moos_Moradi_2021}.
Thus, we initially explore the role of Vt cell selection for the state registers for an AES crypto core under S-PSCA (\autoref{fig:base}).
Note that all setup details are provided in Sec.~\ref{sec:setup}.

From \autoref{fig:base}, we note two key observations.
First, the role of LVT cells (versus HVT and RVT cells) is significant. The more LVT cells are used for the sensitive state registers, the less effort is required for a successful S-PSCA. 
Note here that the assignment of LVT cells to state registers is randomized, to avoid any bias for such circuit-level details, and the results are averaged across multiple runs. Second, the S-PSC resilience is relatively weak as less than 10K traces are sufficient.

\begin{figure}[t]
    \centering
    \includegraphics[width=0.99\columnwidth]{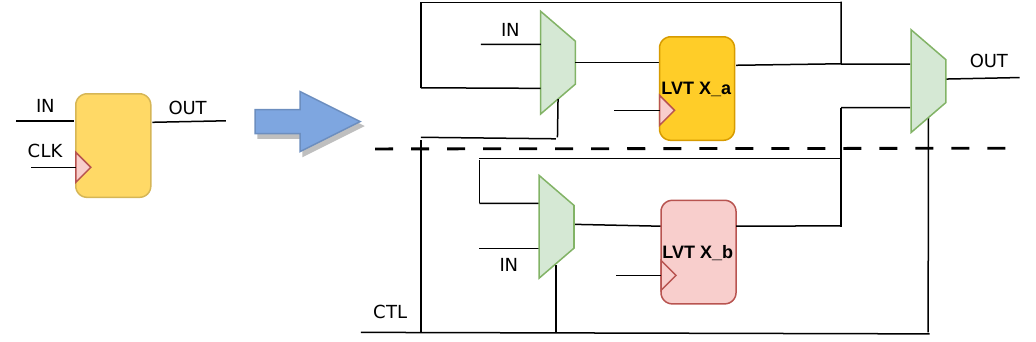}
    \caption{Circuitry primitive for the proposed lightweight countermeasure scheme. The control signal (CTL) is connected to a random binary number generator. %(not illustrated).
	    The dashed line indicates the option of using only the lower path.
    }
    \label{fig:circuit}
    \vspace{-5mm}
\end{figure}

\subsection{Proposed Defense Against S-PSCA}

% Instead of directly linking sensitive data into state registers, we propose an extended circuit primitive for those registers as illustrated in \autoref{fig:circuit}, where the original FF (left) is transformed into a circuit primitive (right). The randomized changes in power patterns introduced by our scheme hide/mask the actual power consumption, making it harder for attackers to understand or predict them using their usual methods.

% More specifically, a regular state register is extended into an entangled implementation of LVT FFs of different driver strengths.
% Note the following for the primitive's working and the resulting resilience against S-PSCAs.
Instead of directly linking sensitive data to state registers, we propose an extended circuit primitive, as shown in \autoref{fig:circuit}. This transforms the original flip-flop (FF) into a more secure form by introducing randomized power pattern changes, masking actual power consumption and making it harder to attack.
Specifically, a regular state register is extended into an entangled implementation of LVT FFs with varying driver strengths, enhancing resilience against S-PSCAs.

First, the primitive could be tailored for any number of paths with different driver-strength LVT FFs. In this work, we consider only 1 or 2 paths, as also indicated by the dashed line in the figure.
These options already increase the resilience of the cipher core significantly (Sec.~\ref{sec:exp}).

% Second, we utilize LVT FFs first and foremost for the fact that their static power is significantly larger and more varied for different data conditions than RVT and HVT cells (\autoref{tab:DFF_leak}). We also utilize
% LVT FFs for their fast switching behaviour: given that the primitive incorporates additional components, the timing delay on the paths covered by these primitives increases, but with the use of LVT FFs this impact is limited. Thus, timing closure becomes easier again.
Second, we use LVT FFs primarily because their static power is significantly larger and more varied under different data conditions compared to other cells (\autoref{tab:DFF_leak}). Their fast switching behavior mitigates the timing delay introduced by the added components, making timing closure easier.

% Third, for the default configuration of the primitive shown in \autoref{fig:circuit}, upon selection of `0' for the control signal (CTL), the primitive operates on the upper path with an LVT FF of strength \textit{a}.
% Conversely, selecting `1' activates the lower path with an LVT FF of strength \textit{b}.
% It is important to understand that both paths produce functionally equivalent outputs; there is no disruption of any of the regular cipher operations whatsoever.
% Rather, these paths are masked variants of each other,
% distinguished only by their Vt cells and driver strengths and their different impact on leakage power.
Third, for the default configuration shown in \autoref{fig:circuit}, selecting '0' for the control signal (CTL) activates the upper path with an LVT FF of strength \textit{a}, while selecting '1' activates the lower path with an LVT FF of strength \textit{b}. Both paths produce functionally equivalent outputs, ensuring no disruption to regular cipher operations. These paths are masked variants, distinguished only by their Vt cells, driver strengths, and their different impact on leakage power.

Finally, and most important for security, the randomized switching between the different paths introduces a layer of unpredictability into the S-PSC patterns as follows.
Note the data feedback loops from the FFs' outputs back to their inputs, which are only active when the corresponding path is inactive.
Thus, the path that is not active at any given cycle will express data-dependent S-PSC information leakage based on the \textit{previous} cycle which may or may not align with the data in the current cycle.\footnote{%
% Here it is important to note that, while S-PSCAs do not require the IC to be running during the actual power measurement, an attacker would still need to activate the clock intermittently,
% namely to allow for the different rounds of AES to be processed in particular and for different cipher messages to be processed in general,
% all to gather power traces for different (static) data hold in the sensitive FFs.
% Naturally, the toggling of the clock signal could be utilized -- in a randomized manner -- for the CTL signal.
While S-PSCAs do not require the IC to be running during measurement, an attacker must intermittently activate the clock to process different AES rounds and cipher messages, gathering power traces for static data in
sensitive FFs. That clock signal toggling can be used randomly for the CTL signal.}
Furthermore, given the exploratory study in \autoref{fig:base}, the application of the circuit primitive should be tuned carefully.
As we find during our experiments (Sec.~\ref{sec:exp}), it is not about limiting/reducing the number of primitives in general
%(as one might guess from \autoref{fig:base} where more LVT cells lead to lower resilience)
but rather about a wide range for LVT versus RVT/HVT cells across the different bytes of the state registers.

In short, there are two-fold noise patterns arising for the S-PSC thanks to our proposed defense:
1)~for the system-level S-PSC -- which is also the one encountered by real-world attackers -- the varying
instantiation of the primitive across bytes incurs considerable diverse power patterns (due to LVT cells in the primitive versus RVT/HVT cells in regular FFs);
2)~for the byte-level S-PSC -- which naturally contributes to the system-level S-PSC -- the randomized feeding of current-cycle versus previous-cycle data to the different-driver strength FFs.
These noise patterns significantly increase the number of samples needed for a successful S-PSCA, drastically raising the attack complexity, as shown in Sec.~\ref{sec:exp}.
% Due to the obfuscation of power traces by these noise patterns, a substantially greater number of samples would be required to perform a successful S-PSCA, drastically increasing the complexity and effort needed for any such
% attempt. This augmentation in security is shown in Sec.~\ref{sec:exp}, where we provide a comprehensive analysis.

\subsection{Implementation Details}
\label{sec:details}

\begin{figure}[t]
    \centering
    \includegraphics[width=\columnwidth, height=4cm]{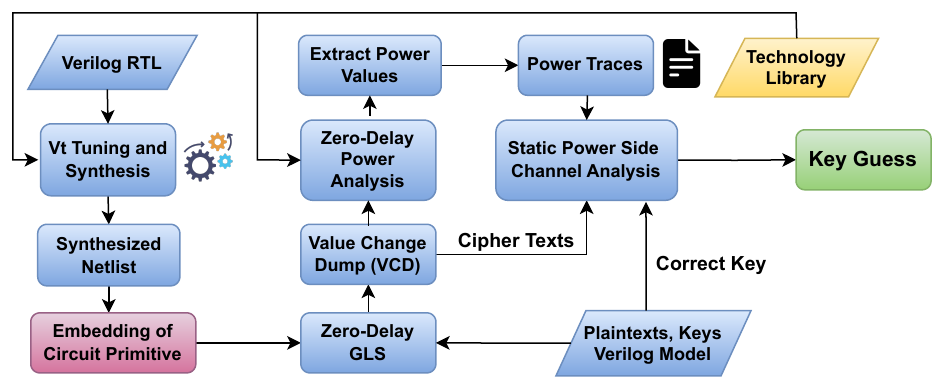}
    \caption{Flow diagram for implementation as well as assessment of our lightweight masking scheme.
	    Note that GLS is short for gate-level simulation.
    }
   
    \label{fig:flow}
   % \vspace{-6mm}
\end{figure}

\autoref{fig:flow} outlines the workflow for our scheme. Details are discussed next.

\subsubsection{Design Implementation}
Initially, RTL of the crypto core (AES in this work) is synthesized using the chosen technology libraries, taking into account all available Vt cell types.
Next, the proposed countermeasure circuitry  is embedded as in replacing all instances of state registers.
Next, the netlist's functionality, post-synthesis, is confirmed by running the testbench. The testbench has a series of plain-texts alongside a key or multiple keys. Subsequently, the testbench generates the cipher-texts which are then validated against a golden software implementation of the crypto core.

\subsubsection{Simulation-Based Power Analysis}
\label{sec:zero-delay_flow}

During gate-level simulations, a value change dump (VCD) file captures data patterns across all gates/nodes at a user-defined temporal resolution, such as $1$~ps. This VCD file, along with the post-synthesis netlist and selected libraries, is used for power simulation, evaluating 1) static/leakage power, 2) internal power from input-pin switching, and 3) switching power from output-pin switching. We focus on static power using zero-delay simulations to capture specific clock cycles, as static power values remain constant regardless of simulation type. Power traces are obtained sequentially while processing different texts for the specified secret keys, focusing on final round operations as outlined in \cite{brier2004CPA}.

\subsubsection{Sampling-Based CPA Attack}
\label{sec:our_CPA}

CPA uses the Pearson correlation coefficient (PCC) to link power consumption during cryptographic operations to confidential data~\cite{brier2004CPA}. This involves comparing actual power usage with expected patterns for different key values, deducing the key byte by byte from the highest PCC values.
In S-PSCA, expected power patterns are based on either the Hamming weight (HW) of ciphertexts or the Hamming distance (HD) between ciphertexts and the final-round operation, depending on the technology node~\cite{brier2004CPA}. Real attackers use both models.
We enhance CPA with a sampling framework for robust insights, iteratively increasing the number of traces and tracking success rates. The process stops once a desired confidence level, like 90\%, is achieved.
For efficiency, we use multi-threading and two-phase sampling: coarse sampling with fewer permutations and larger steps to find a starting point, followed by detailed sampling with more trials and smaller steps.

\section{Experiments\label{sec:exp}}

\subsection{Setup}
\label{sec:setup}
We conduct all experiments, including the preliminary analysis in Sec.~\ref{sec:framework}, on an AMD EPYC 7542 server with Red Hat Enterprise Linux Server. For our CAD flow, we use various commercial tools: {Synopsys
	VCS M-2017.03-SP1} for both RTL and gate-level functional simulations, {Synopsys DC M-2016.12-SP2} for logic synthesis, and {Synopsys PrimeTime PX M-2017.06} for power simulations. The sampling-based CPA
	framework is implemented upon the open-source code in \cite{knechtel22_PSC}.
% We conduct all experiments, including the preliminary analysis presented in Sec.~\ref{sec:framework}, on an AMD EPYC 7542 server
% with Red Hat Enterprise Linux Server Release 7.9.
% In terms of the CAD flow for our implementation, we utilize various commercial tools. For both RTL and gate-level functional simulations, we employ {Synopsys
% 	VCS M-2017.03-SP1}. Logic synthesis is carried out using {Synopsys DC M-2016.12-SP2}, while power simulations are performed with {Synopsys PrimeTime PX M-2017.06}.
% The sampling-based CPA framework is implemented upon the open-source code in \cite{CPA_orig}.
	% We employ a commercial library for the 28nm node. We consider the TT corners which are characterized for 25 degrees Celsius and 0.9V.
	% The timing constraint is 0.5n.
	% Standard optimization techniques are applied
	% during the synthesis process, which implies that LVT cells are used sparingly in the baseline designs. In fact, none of the state registers were implemented using LVT cells by default.
	% After integration of the proposed circuit primitive, we apply ECO fixes for timing closure as needed, while making sure that the primitive instances themselves are not revised/optimized.
	% The overheads resulting from ECO fixing are included in cost reports.
 We use a commercial 28nm library with TT corners characterized at 25 degrees Celsius and 0.9V, applying standard optimization techniques sparingly for LVT cells in the baseline designs. 
 %None of the state registers use LVT cells by default.
 After integrating the proposed circuit primitive, we apply ECO fixes for timing closure without revising or optimizing the primitive instances, and the resulting overheads are included in cost reports.

For the AES design, we leverage a regular RTL that is working on 128-bit keys and 128-bit texts, uses look-up tables for the S-Box, and has no other PSC countermeasure in place.

% For all CPA runs, we consider 1M traces in total. We employ coarse versus thorough sampling considering 64 versus 640 trials, respectively,
%     and we report final results of $t$ traces until disclosure for a 90\% success rate for thorough sampling, i.e., CPA must succeed to infer all key bytes correctly for at least 576 out of 640 randomly selected subsets of $t$ traces.
% We consider HW and HD models, and find that the HD model has better results in all cases; thus, final results are reported for the HD model.
For all CPA runs, we use 1 million traces in total, with coarse sampling of 64 trials and thorough sampling of 640 trials. We report the number of traces until disclosure (NTTD) for a 90\% success rate, where CPA must correctly infer all key bytes in at least 576 out of 640 trials. Both HW and HD models are considered, but the HD model performs better in all cases, so final results are reported for the HD model.

%%%%%%%%%%%%%%%%%%%%%%%%%%%%%%%%%%%%%%%%%%%%%%%%%%%%%%%%%

%%%%%%%%%%%%%%%%%%%%%%%%%%%%%%%%%%%%%%%%%%%%%%%%%%%%%%%%%%%%%%%%%%%%%%%%%%%%%%%%%%

%%%%%%%%%%%%%%%%%%%%%%%%%%%%%%%%%%%%%%%%%%%%%%%%%%%%%%%%%%%%%%%%%%%%%%%%%%%%%%%%%%%%%%%
% \begin{table*}[]
% \caption{Previous Work for S-PSCA Countermeasures in PRESENT Core \cite{Moos_Moradi_2021}.}
% \label{tab:results1}
% \centering
% \scriptsize
% \begin{tabular}{|c|c|c|c|c|}
% \hline
% \multirow{2}{*}{Design } & \multirow{2}{*}{ Area } & \multirow{2}{*}{ Overhead } & \multirow{2}{*}{ \# of Power Traces (PT) } & \multirow{2}{*}{ PT / Area } \\ 
% & & & & \\ \hline
% High Performance (HP) & 2 535.00 & ×1.00 & $<$ 100 & $<$ 0.04 \\ \hline
% High Threshold Voltage (HVT) & 2 406.67 & ×0.95 & 200 & 0.08 \\ \hline
% Exhaustive Logic Balancing (ELB) & 20 207.00 & ×7.97 & 120 000 & 5.94 \\ \hline
% %& & & & \\
% Threshold Implementation + HP & 7 233.33 & ×2.85 & 23 600 & 3.26 \\ \hline
% Threshold Implementation + HVT & 6 982.67 & ×2.75 & 53 000 & 7.59 \\ \hline
% Threshold Implementation + ELB & 58 442.33 & ×23.05 & 2 930 000 & 50.14 \\ \hline
% \end{tabular}

% \end{table*}
%%%%%%%%%%%%%%%%%%%%%%%%%%%%%%%%%%%%%%%%%%%%%%%%%%%%%%%%%%%%%%%%%%%%%%%%%%%%%%%%

\begin{table}[b]
\caption{CPA Results for LVT Tuning Only}
\label{tab:results2}
\centering
\scriptsize
\begin{tabular}{|c|c|c|c|c|c|c|}
\hline
& & 0 & 2 & 4 & 6 & 8 \\ \hline
\multirow{5}{*}{ Dataset 1 } & 0 & --- & --- & --- & --- & --- \\
& 2 & 147 000 & --- & --- & --- & --- \\
& 4 & 378 000 & 80 000 & --- & --- & --- \\
& 6 & 610 000 & 221 000 & 43 000 & --- & --- \\
& 8 & 615 000 & 321 000 & 140 000 & 31 000 & --- \\ \hline
%& & & & & & \\
\multirow{5}{*}{ Dataset 2 } & 0 & --- & --- & --- & --- & --- \\
& 2 & 154 000 & --- & --- & --- & --- \\
& 4 & 401 000 & 71 000 & --- & --- & --- \\
& 6 & 518 000 & 191 000 & 42 000 & --- & --- \\
& 8 & 650 000 & 348 000 & 130 000 & 29 000 & --- \\ \hline
%& & & & & & \\
\multirow{5}{*}{ Dataset 3 } & 0 & --- & --- & --- & --- & --- \\
& 2 & 160 000 & --- & --- & --- & --- \\
& 4 & 401 000 & 73 000 & --- & --- & --- \\
& 6 & 420 000 & 240 000 & 41 000 & --- & --- \\
& 8 & 809 000 & 381 000 & 132 000 & 31 000 & --- \\ \hline
%& & & & & & \\
\multirow{5}{*}{ Average } & 0 & --- & --- & --- & --- & --- \\
& 2 & 153 667 & --- & --- & --- & --- \\
& 4 & 393 333 & 74 667 & --- & --- & --- \\
& 6 & 516 000 & 217 333 & 42 000 & --- & --- \\
& 8 & \textbf{691 333} & 350 000 & 134 000 & 30 333 & --- \\ \hline
\end{tabular}\\[1mm]
Rows indicate the number of primitives for one half of the state registers bytes, and columns indicate the same for the other half.
\end{table}

%%%%%%%%%%%%%%%%%%%%%%%%%%%%%%%%%%%%%%%%%%%%%%%%%%%%%%%%%%%%%%%%%%%%%%%%%%%%%%%
\begin{table}[t]
\caption{CPA Results for LVT and Driver-Strength Tuning}
\label{tab:results3}
\centering
\setlength{\tabcolsep}{5pt}
\scriptsize
\begin{tabular}{|c|c|c|c|c|c|c|}
\hline
& & 0 & 2 & 4 & 6 & 8 \\ \hline
\multirow{5}{*}{ Dataset 4 } & 0 & 19 000 & 161 000 & 403 000 & 884 000 & 612 000 \\
& 2 & 113 000 & 12 000 & 105 000 & 260 000 & 421 000 \\
& 4 & 270 000 & 31 000 & 12 000 & 80 000 & 196 000 \\
& 6 & 420 000 & 102 000 & 10 000 & 15 000 & 72 000 \\
& 8 & 467 000 & 188 000 & 51 000 & 4 000 & 17 000 \\ \hline
%& & & & & & \\
\multirow{5}{*}{ Dataset 5 } & 0 & 18 000 & 171 000 & 334 000 & 487 000 & 924 000 \\
& 2 & 121 000 & 13 000 & 114 000 & 287 000 & 378 000 \\
& 4 & 241 000 & 31 000 & 13 000 & 82 000 & 190 000 \\
& 6 & 340 000 & 111 000 & 11 000 & 15 000 & 70 000 \\
& 8 & 500 000 & 200 000 & 50 000 & 4 000 & 18 000 \\ \hline
%& & & & & & \\
\multirow{5}{*}{ Dataset 6 } & 0 & 21 000 & 173 000 & 327 000 & 723 000 & 998 000 \\
& 2 & 114 000 & 13 000 & 102 000 & 230 000 & 371 000 \\
& 4 & 240 000 & 30 000 & 13 000 & 79 000 & 202 000 \\
& 6 & 338 000 & 109 000 & 11 000 & 14 000 & 70 000 \\
& 8 & 539 000 & 181 000 & 52 000 & 4 000 & 16 000 \\ \hline
%& & & & & & \\
\multirow{5}{*}{ Average } & 0 & 19 333 & 168 333 & 354 667 & 698 000 & \textbf{844 667} \\
& 2 & 116 000 & 12 667 & 107 000 & 259 000 & 390 000 \\
& 4 & 250 333 & 30 667 & 12 667 & 80 333 & 196 000 \\
& 6 & 366 000 & 107 333 & 10 667 & 14 667 & 70 667 \\
& 8 & 502 000 & 189 667 & 51 000 & 4 000 & 17 000 \\ \hline
\end{tabular}\\[1mm]
Rows indicate the number of primitives for one half of the state registers bytes, and columns indicate the same for the other half.
For the first half, all FFS have the same driver strength (randomly selected once), whereas for the second half, driver strengths are randomly assigned across all FFs.

\end{table}

\subsection{Results I: LVT Tuning Only}

We used the AES baseline design and incorporated circuit primitives as shown in \autoref{fig:circuit}, utilizing only one path. Primitives were instantiated with the same driver strength across all state registers to meet worst-case timing closure needs. For the 16 bytes of state registers, we varied the number of primitives, generating multiple design sets with lightweight countermeasures.
For each set, we randomly selected 8 bytes and assigned 0, 2, 4, 6, or 8 primitives per byte at random bit-level positions. The remaining 8 bytes had a constant number of primitives for each set. This experiment aimed to explore the search space for varying numbers of instances in different parts of the state registers.

CPA results, as in NTTD for key recovery with 90\% success rate, are presented in \autoref{tab:results2}. Each dataset represents an independently randomized run for design generation.
Rows indicate the number of primitives for one half of the state registers bytes, and columns indicate the same for the other half.
For example, the intersection of row 4 and column 2 represents a design where four primitives per byte are used in one half and two primitives per byte are used in the other half. Empty cells (---) represent scenarios that are either not applicable or duplicates.
We generated multiple datasets and found consistent trends across three different datasets, leading us to average the results. The outcome is significant: up to \textbf{79 times} more power traces were needed with our defense mechanisms compared to the most resilient baseline design with simple LVT assignment (\autoref{fig:base}).

The largest variation of 0 versus 8 primitives induced the highest resilience. This can be explained by the inherent limitation of CPA, which decomposes the problem at the byte level into 16 bytes with $2^8$ possible candidates each. Each byte is attacked separately, but the remaining 15 bytes still contribute to the system-level PSC observable by attackers. This attack method inherently deals with noise patterns. Regular IC implementations, including prior masking methods, exhibit relatively small variations in power patterns across bytes. Our lightweight countermeasure breaks this premise by inducing significantly larger variations in power patterns.
The area overhead was only up to \textbf{1.02 times}  of the baseline design, demonstrating the efficiency of our approach. The results are summarized in \autoref{tab:results_our}, in \textbf{Countermeasure-I (LVT).}

\subsection{Results II: LVT and Driver-Strength Tuning}

Here we extend our study to tuning of both LVT and driver strengths.
The process for design generation is very similar to the previous experiment, except for the following.
We randomly assign varying driver strengths for all FFs (within or outside of primitives) of the second half of the state-register bytes, whereas all FFs of the first half remain at constant driver strength, which is still randomly
selected.
For varying driver strengths within primitives, here we also consider the full circuit primitive shown in \autoref{fig:circuit}, i.e., with two paths. 
In general, driver strengths are selected from the range of X2, X4, and X8.\footnote{Strength X2 is sufficient for timing closure, as the LVT FF only has to drive one MUX; higher strengths are purely utilized for S-PSC obfuscation.}
The intent of this experiment is to further enhance the variability of the S-PSC.

CPA results are shown in \autoref{tab:results3}. This table follows the same general structure as \autoref{tab:results2}, but, since columns represent the second half of the bytes which are randomly tuned for
driver strength now, all entries are meaningful. For example, row `4' and column `4' now cover design cases where four primitives per byte (each with one path and constant but randomly selected driver
strength for its LVT FF) are used in one half and four primitives per byte (each with independently randomly selected driver strengths for its two LVT FFs) are used in the other half of state registers.

The outcome here is even more significant: on average, up to \textbf{96 times} more power traces were needed when compared to the most resilient case for the baseline design
(\autoref{fig:base}). 
The largest variations in terms of primitives counts again induced the highest resilience. In more detail, we find that random variations of driver strengths for all 8 primitives per
byte for one half (along with no primitives but regular RVT/HVT FFS with constant-but-random strengths for the other half), is more effective -- 844 667 traces are needed on average for a successful attack -- than random
driver-strength variations for 8 RVT/HVT FFs per byte (along with 8 primitives with constant-but-random strengths for their single LVT FF) where 502 000 traces are needed.
In short, tuning the driver strengths in full-scale primitives is more impactful.

The area overhead was still not more than {1.06 times} that of the baseline design. The marginally larger overhead here is due to the use of two paths for the circuit primitive. Still, when putting the
NTTD into the design context (i.e., traces over area), the configuration covered in this experiment even more efficient.
Related results
are summarized in \autoref{tab:results_our}, in \textbf{Countermeasure-II (LVT + Driver Strength)}.

\begin{table}[]
\caption{S-PSCA Countermeasures in AES Core}
\label{tab:results_our}
\centering
\setlength{\tabcolsep}{3pt}
\scriptsize
\begin{tabular}{|c|c|c|c|c|} 
\hline
\multirow{2}{*}{ Design} & \multirow{2}{*}{ Area [$\mu$m$^2$]} & \multirow{2}{*}{ Overhead } & \multirow{2}{*}{ NTTD } & \multirow{2}{*}{ NTTD / Area } \\ 
& & & & \\ \hline
Our Baseline  & 13 231.39 & x1.00 & 8 780 &0.66 \\  \hline
Our Countermeaure-I & \multirow{2}{*}{13 503.15} & \multirow{2}{*}{x1.02} & \multirow{2}{*}{691 333} & \multirow{2}{*}{51.19} \\ 
(LVT) &  &  &  &  \\ \hline
Our Countermeaure-II  & \multirow{2}{*}{14 021.41} & \multirow{2}{*}{x1.06} & \multirow{2}{*}{844 667} & \multirow{2}{*}{\textbf{60.24}} \\ 
(LVT + Driver Strength) &  &  &  &  \\ \hline
\hline
ELB\cite{ASCENT} & 25 928 & x1.97 & 87 000 & 3.35 \\ \hline
%\cite{dumitru2023borrowedtimepreventing}  &   - & - & 1 000 000 & - \\ \hline
QuadSeal~\cite{QuadSeal} & 131 435 & x6.5 &  1 000 000 & 7.61 \\ \hline
\cite{ML-assisted_countermeasure} &  230 000    & x1.36 & 1 500 000 & 6.52\\ \hline
\end{tabular}
\end{table}

%%%%%%%%%%%%%%%%%%%%%%%%%%%%%%%%%%%%%%%%%%%%%%%%%%%%%%%%%%%%%%%%%%%%%%%%%%%%%%%%%%%%%%%%%%%%%%
\subsection{Comparison to Prior Art}

% The most recent and relevant prior work in S-PSCAs is given in \cite{Moos_Moradi_2021}; also recall Sec.~\ref{sec:related}. Their key results are quoted in \autoref{tab:results1}. Although their work utilizes not AES but PRESENT, a high-level comparison is still interesting, as both ciphers are S-Box-based and have very similar operation principles. The joint metric for effectiveness and efficiency (traces over area; the higher the better) indicates that ours is competitive to their work.
% Since our study is based on simulated (but technology accurate) power data, unlike theirs based on measurement data, we can expect that metric to scale even further up for ours, as more traces will be needed for real-world attackers that face measurement noise.

%The most recent and relevant prior work in S-PSCAs is given in \cite{our_paper1, Moos_Moradi_2021}; also recall Sec.~\ref{sec:related}. Their key results are quoted in \autoref{tab:results1}, we have put (---) for the information missing in the respective papers.
% Although their work utilizes not AES but PRESENT, a high-level comparison is still interesting, as both ciphers are S-Box-based and have very similar operation principles. 

In \autoref{tab:results_our}, we also compare to relevant prior art, i.e., those that consider AES and S-PSCA evaluation.
In terms of NTTD over design area, i.e., in terms of efficiency, ours is far superior compared to those works.
Importantly, since our work is based on technology-accurate simulated power data whereas others on measured power data, we can expect the efficiency of ours to scale even further up (as more traces will be needed for
		real-world attackers that face measurement noise).

\begin{comment}

%%%%%%%%%%%%%%%%%%%%%%%%%%%%%%%%%%%%%%%%%%%%
\begin{table}[]
\caption{Previous Work for PSCA Countermeasures in AES Core}
\label{tab:results1}
\centering
\scriptsize
\begin{tabular}{|c|c|c|c|c|}
\hline
\multirow{2}{*}{Design } & \multirow{2}{*}{ Area [$\mu$m$^2$]} & \multirow{2}{*}{ Overhead } & \multirow{2}{*}{ NTTD } & \multirow{2}{*}{ NTTD / Area } \\ 
& & & & \\ \hline
% High Performance (HP) & 2 535.00 & ×1.00 & $<$ 100 & $<$ 0.04 \\ \hline
% High Threshold Voltage (HVT) & 2 406.67 & ×0.95 & 200 & 0.08 \\ \hline
% Exhaustive Logic Balancing (ELB) & 20 207.00 & ×7.97 & 120 000 & 5.94 \\ \hline
% %& & & & \\
% Threshold Implementation + HP & 7 233.33 & ×2.85 & 23 600 & 3.26 \\ \hline
% Threshold Implementation + HVT & 6 982.67 & ×2.75 & 53 000 & 7.59 \\ \hline
% Threshold Implementation + ELB & 58 442.33 & ×23.05 & 2 930 000 & 50.14 \\ 
%\hline
%\cite{extremeley_light_dpsc} (DPSC)  & - &  x1.04 & 50 000 & - \\ \hline
ELB\cite{ASCENT} & 25 928 & x1.97 & 87 000 & 3.35 \\ \hline
%\cite{dumitru2023borrowedtimepreventing}  &   - & - & 1 000 000 & - \\ \hline
QuadSeal~\cite{QuadSeal} & 131 435 & x6.5 &  1 000 000 & 7.61 \\ \hline
\cite{ML-assisted_countermeasure} &  230 000    & x1.36 & 1 500 000 & 6.52\\ \hline

\end{tabular}

\end{table}

\end{comment}

\section{Conclusion\label{sec:conclusions}}
% In this work, we have proposed lightweight countermeasure as defense against S-PSCAs. Our proposed scheme is based on utilizing largely different threshold-voltage cells and driver strengths across the bytes of the sensitive state registers of a crypto core. Our approach introduces significant noise patterns in the power profile, at multiple levels, leading to higher resilience which is confirmed through robust, sampling-based, CPA attack campaigns.
% Our approach incurs minimal overhead which is quite in contrast to prior art: we incur only \textbf{1.06 times} of overhead, all while imposing up to \textbf{96 times} higher resilience. The corresponding `traces over area' metric puts ours in good competition with prior art.

% In future work, we will further validate our scheme via measurement campaigns and extend the implementation to other crypto cores. Both aspects represent `only' engineering challenges but not fundamental ones. Thus, our scheme
% is expected to excel even further for these broader use cases.

In this work, we propose a lightweight countermeasure against S-PSCAs. Our scheme uses varied threshold-voltage cells and driver strengths across sensitive state registers in a crypto core, introducing significant noise
in the power profile and enhancing resilience. Robust CPA attack campaigns confirm the effectiveness of our approach. Despite the minimal overhead of only \textbf{1.06 times}, our method offers up to \textbf{96 times}
higher resilience. The ``traces over area'' metric demonstrates superior performance over prior art.
Future work will further validate our scheme through measurement campaigns and extend it to other crypto cores.
%\footnote{%
%	As this study is based on technology-accurate simulated power data, we can expect the performance of our scheme to scale even further up, as more traces will be needed for real-world attackers that face measurement noise.}
Importantly, such aspects are engineering challenges rather than fundamental ones.
%promising further excellence in broader applications.

\bibliographystyle{IEEEtran}
%\scriptsize{
    \bibliography{main.bib}
%}

% \bibliographystyle{IEEEtran}
% \bibliography{biblio}

\end{document}